# Fabrication of p$^+$ contact by Thermally Induced Solid State Regrowth of Al on p-type Ge Crystal

Manoranjan Ghosh,[1,a)] Shreyas Pitale,[1] S.G. Singh,[1] Husain Manasawala,[1]
Vijay Karki,[2] Manish Singh,[2] Kulwant Singh,[3] G. D. Patra,[1] Shashwati Sen[1,4]

*Abstract*—Formation of p$^+$ contact on Germanium is important for applications in diode detector and other electronic devices. In this work, thermally deposited Al on Ge crystal is annealed at 350ºC followed by slow cooling for solid-state regrowth of Al-Ge p$^+$ contact on Ge. Depth profile analysis by secondary ion mass spectrometry (SIMS) is carried out to investigate the occurrence of Al and Ge along the depth of the regrown Al-Ge layer. Evidence of regrowth is observed due to inter-diffusion of both Ge and Al across the layers although Ge diffusion in Al layer is found to be more prevalent. Thickness of the evaporated Al layer is varied to understand the diffusion profile of Al, Ge and estimate the depth of Al incorporation in Ge crystal underneath. Hall measurement at different depth of Al-Ge regrown layer reveals that Al impurity induces p$^+$ doping in p-type Ge and its concentration gradually reduces towards the Ge crystal. Top surface of the Al-Ge layer exhibits lowest sheet resistance that varies with the thickness of the as deposited Al layer.

## I. Introduction

Owing to its high carrier mobility and efficient absorption for gamma-rays (Z=32) Germanium diode based detectors are coveted in gamma ray spectroscopy for their superior resolution and efficiency [1,2]. Although operation of these detectors requires cryogenic cooling, the typical resolution of high purity germanium (HPGe) detectors is about 0.2% at 662 keV, compared with that of NaI scintillators which is typically ~7% [3]. Suitable electrical contacts are required for application of Ge as diode detector. Planar detectors are fabricated by forming electron and hole blocking contacts on opposite faces of intrinsic Ge crystal [4]. Conventionally, lithium diffusion is used to fabricate the n$^+$ contact [5] and boron implantation for p$^+$ contact [6]. Since extensive and sophisticated procedure is involved for boron ion implantation, alternative contacts were explored [7,8].

In order to identify appropriate contact materials, thermally induced reaction of 20 transition metals with Ge substrates has been studied. Phase formation of each metal-Ge system has been determined for the identification of the most promising candidates—in terms of sheet resistance and surface roughness [9]. To avoid further addition of impurities, low temperature reactions have been employed to form thin blocking or ohmic contacts to semiconductors such as Si and GaAs [10,11]. It is found that *n-* and *p-type* layers can be formed in Ge at temperatures below 300$^0$C by using solid-solid reactions between Ge and evaporated metal films [4]. Epitaxial growth of Ge is claimed from a solid solution of Ge in an Al film onto n-type Ge substrates. Hall voltage polarities and hot point probe measurements indicated p-type conductivity of the grown layer [12]. In a similar study, solid Al was used as a medium from which to grow Ge onto a crystalline Ge substrate and the regrown layer is found to be *p* –type [13]. Ge alpha particle detector has been fabricated by employing solid-phase epitaxial regrowth technique to grow p-type contacts on n-type Ge by evaporating 40-100 nm Al layer followed by annealing and slow cooling [14]. Further, HPGe detector has been fabricated using thin p+ contacts formed by solid phase regrowth of evaporated Al layers on Ge at 300°C and lithium diffused n+ contacts [15].

Al-Ge contacts are epitaxial in nature when grown below its eutectic temperature (424$^0$C) and exhibits p-type conductivity as established by all earlier studies. Mainly, channeling effect and backscattering measurements were performed to determine the orientation and crystalline nature of regrown Ge [12]. In this work, Al-Ge solid-state regrown p$^+$ contact is fabricated on Ge by thermal deposition of Al on Ge followed by heat treatment at 350ºC. Depth profile analysis by secondary ion mass spectrometry (SIMS) is performed to investigate the presence of Ge and Al along the depth of regrown layer. Also, sheet resistance, impurity concentration and type of conductivity have been determined by hall measurement along the depth of regrown layer for its applicability as p+ contacts on Ge. Thickness of the deposited Al layer has been varied systematically and its effect on the length of regrown layer as well as diffusion profile of both Ge and Al has been determined which was not studied earlier.

## II. Methods of Sample Preparation and Experimental Techniques

Commercially available Ge crystals (carrier conc. ~10$^{10}$/cc at 77K) were cut by diamond wheel to a suitable size and shape. Pieces having flat faces and desired size of 8mm × 8mm × 1.5 mm were achieved by lapping on SiC abrasives sheets successively with grit sizes from 220 to 1500. Samples were cleaned by ultra-sonication in methanol for 15 minutes. Further planarization is performed and mirror like surfaces are obtained by polish etching in 3:1 solution of $HNO_3$:HF (by volume) for 3 minutes. The substrates were then mounted in a thermal evaporation system equipped with provision of in-situ annealing in argon environment after deposition. Al films of various thicknesses were deposited on Ge using 5-20 mg of Al wire wound around the tungsten filament fastened between the electrodes. The chamber was pumped down to $2\times10^{-6}$ mbar using a turbo-molecular pump and the current through the filament was gradually increased up to about 50A to evaporate Al wire completely.

[1]*Technical Physics Division, Bhabha Atomic Research Centre*
[2]*Fuel Chemistry Division, Bhabha Atomic Research Centre*
[3]*Material Science Division, Bhabha Atomic Research Centre*
[4]*Homi Bhabha National Institute*
[a)]Email: mghosh@barc.gov.in

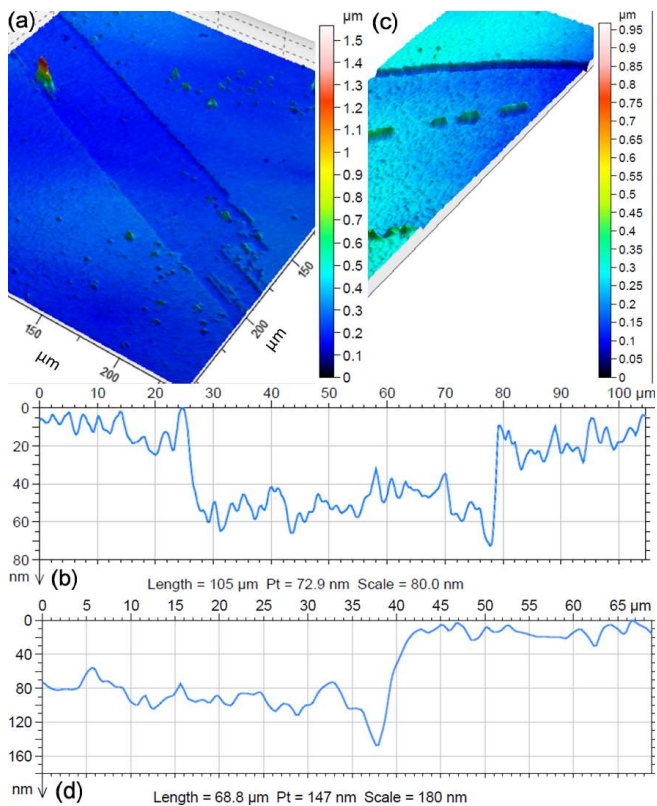

Fig. 1. Surface profile (a) and thickness (b) of scratched Al film on Ge after thermal deposition of 5 mg Al without post deposition heat treatment. The same for 15 mg Al is shown in (c) and (d).

TABLE I
THICKNESS OF DEPOSITED AL FILM FOR VARIOUS AMOUNTS OF SOURCE MATERIALS (AL)

| Mass of Al | Film Thickness |
|---|---|
| 5 mg | ~30 nm |
| 10 mg | ~60 nm |
| 15 mg | ~90 nm |
| 20 mg | ~120 nm |

For the regrowth of Ge in the as deposited Al layer, annealing of samples was performed immediately after thermal evaporation in an argon atmosphere. The process temperature profile was set and controlled by a PID temperature controller. As deposited Al coated Ge (hence forth Al/Ge) samples were first heated up to $350^0$C within 20 minutes and maintained for 40 minutes before slowly cooled down to $30^0$C in 90 minutes. Evidence of regrowth and Ge diffusion up to the top Al surface can be seen in heat treated Al/Ge samples (hence forth Al-Ge). The easily removable Al layer converts in to rugged p-type Al-Ge contacts on Ge crystal.

Thickness of Al films were determined by Taylor-Hobson make optical 3D profiler (Model-CCI-MP) based on coherence scanning interferometry. Depth profile measurement by SIMS (CAMECA IMS - 7f) is performed to investigate the presence of Al and Ge along the depth of Al-Ge regrown layer. Carrier concentrations, hall coefficient, resistivity and carrier mobility of Al-Ge samples were measured by Ecopia make HMS5000 Hall effect measurement system within a range of temperatures from 77K to 300K. Four contacts were taken on the corners of square sample through In-Ga eutectic in Van der Pauw configuration. The sheet resistance of the samples was measured by collinear 4-probe method at each step of fabrication to qualify the process.

## III. RESULTS AND DISCUSSIONS

### A. Determination of thickness of Al film on Ge

Knowledge of thickness of Al layer on unannealed Al/Ge samples is necessary to investigate the regrown Al-Ge layer. Also measuring the depth of craters created by the impact of ion beam during SIMS measurement is required for finding out the diffusion length of Ge and Al. Thickness measurement have been performed on both as deposited Al layers and craters on Al-Ge regrown layer by 3D optical profilometer. Measured thickness of as deposited Al layers on Ge for different amount of source materials are given in Table 1. Measured crater depths (not shown here) are used for converting the sputtering time in to depth. 3D profile of the Al/Ge surface containing scratches for determining the thickness of the Al film are shown in Fig. 1(a), (b) for 5 mg and (c), (d) for 15 mg Al source material. Height difference between top of the Al film and exposed Ge crystal for 5mg and 15 mg Al are found to be around 30 nm and 90 nm respectively.

Film thickness increases by ~30nm for complete deposition of every 5mg of Al that emerges as an important finding for repetitive deposition of desired thickness of Al film in the current apparatus. Assuming uniform spherical distribution of Al vapour, the thickness (t) can be estimated as $m/4\pi\rho d^2$ where m is the mass of Al, $\rho$ is the density and d is the sample distance from the filament. For, 0.005g Al and the sample being about 82 mm away from the filament (d), the expected thickness is 25 nm which agrees well with the measured thickness by 3D profilometer (Table 1).

### B. Depth profile measurement of Al-Ge regrown layers by SIMS

For making rugged electrical contacts on Ge, Al/Ge samples were annealed at $350^oC$ and slowly cooled as described earlier. The sheet resistance of un-annealed Al/Ge surface is found to be 0.4 ohm and remain unchanged for different Al thickness. Whereas the sheet resistance for annealed Al-Ge layer varies with the thickness of Al layer and found to be 1.5 ohm and 7 ohm for 120 nm and 30 nm Al film. It indicates regrowth of Ge in Al layer because a certain percentage of Ge comes in to super-saturation when the Al-Ge solid solution is cooled down due to difference in solid solubility of Ge in Al at $350^0$C (0.7%) and room temperatures (nil), as observed by prior researcher [16]. SIMS depth profile measurement has been carried out to investigate the occurrence of Al and Ge in the Al-Ge regrown layer. Fig. 2(a) and (b) represents the profile of Al and Ge as a function of depth (converted from known sputtered depth and sputtering time) in annealed and un-annealed sample of 60 nm thick Al layer. The same for 120 nm thick Al layer is shown in Fig. 2(c) and (d). From depth profile analysis of Al and Ge, it is evident that Ge diffuses inside the Al layer and mixing of Ge and Al is observed all through the Al layer in heat treated samples. It indicates reconstruction of Al/Ge interface due to layer exchange and induced crystallization exhibiting significant presence of Ge at the top of Al layer [17,18]. The depth profile of Ge shows preferential growth near the surface (both Ge and Al) because it acts as a growth center for supersaturated Ge [16]. In the contrary, no signature of Ge is found in Al layers in un-annealed Al/Ge samples.

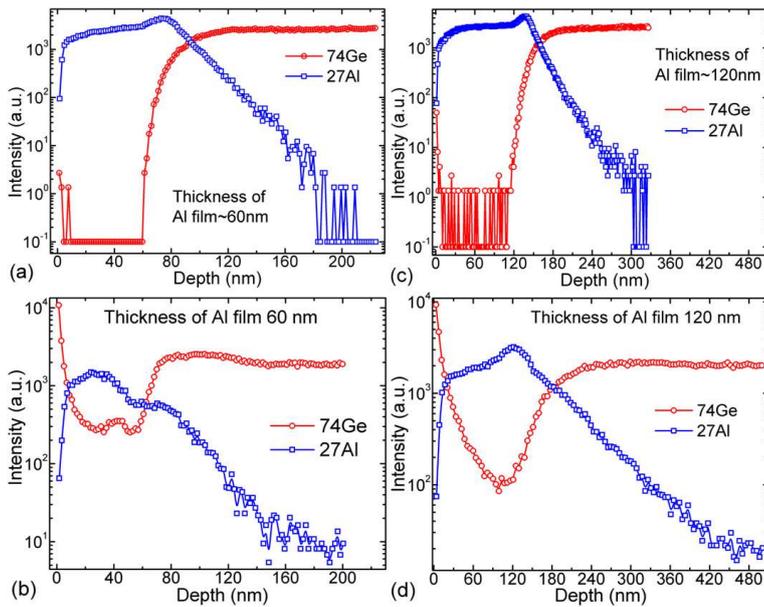

Fig. 2. (a) SIMS profile of Ge and Al along the depth for unannealed Al/Ge [(a) and (c)] and annealed Al-Ge [(b) and (d)] samples for as deposited Al film thickness 60 nm and 120 nm as indicated on the graph.

Evidence of Al diffusion in Ge crystal is also observed in heat treated samples. For 60 nm and 120 nm thick Al deposited un-annealed sample, Al can be found (above 10 counts) up to 60 nm and 120 nm in Ge due to stray Al present in the chamber since Al diffusion in Ge without heating is unlikely. On the other hand, Al can be present up to 200 nm and 480 nm inside Ge in heat treated samples for similar thickness of Al. It indicates Al diffusion in Ge after heat treatment at $350^0$C although Ge diffusion in Al layer is more significant. Thus, the length of Al diffusion in Ge is found to be linked with the as deposited Al film thickness. It is established further by depth profile analysis of two additional samples having Al thickness 15 nm and 150 nm [Fig. 3(a) and (b)]. Diffusion of Al can be seen up to 60 nm and 580 nm inside Ge for Al film thickness around 15 nm and 150 nm respectively. It reaffirms that Al diffusion length in Ge extends up to four times of Al film thickness after heat treatment at $350^0$C.

### C. Hall measurement of Al-Ge samples

Temperature dependent Hall measurement is carried out to investigate the effect of Al incorporation in Ge along the depth of Al-Ge layer. For this purpose, 120 nm Al film is deposited on Ge and regrown at $350^0$C to form around 150 nm Al-Ge layer. Sheet resistance of the top surface of Al-Ge layer is found to be 1.5 ohm (~0.4 ohm for unannealed surface) and exhibits metallic electrical properties. It validates the observed Ge diffusion up to the top surface of Al-Ge layer through SIMS depth profiling. Hall measurement at different depths of Al-Ge layer is performed to probe the existing concentration gradient of Al and its impact on other electrical properties. Al-Ge layers having different depths are created by chemical-mechanical lapping (CML) in five steps for equal duration. Top surface after each CML is characterized by sheet resistance measured by collinear four probes as well as Van der Pauw method. Duration of the CML and layer depth is calibrated using measured sheet resistance values. It was found that approximately 150 nm of Al-Ge layer can be lapped by five steps of CML and sheet resistance values can be used to identify the exposed layers to rule out any possible error in estimating the depth. Temperature dependent sheet resistance of all exposed layers after successive lapping has been plotted in Fig. 4(a) and corresponding estimated depths are also indicated. Top surface of Al-Ge layer without lapping exhibits metallic character and discussed separately (Fig. 5). After first stage of lapping, the sheet resistance at room temperature increases from 1.5 ohm to 15 ohm. It indicates that Al-Ge at this depth contain high level of electrically active Al impurity which is reflected by reduction in sheet resistance with decrease in temperature. With successive lapping of Al-Ge layer, there is consistent increase in sheet resistance up to 600 ohm at a depth of around 150 nm. Further lapping exposed the pure germanium surface. It indicates that Al concentration gradually reduces as the depth of Al-Ge layer increases. Also the Al-Ge layer beyond 60 nm depth exhibits semiconductor type behavior and the sheet resistance increases with the decrease in temperature.

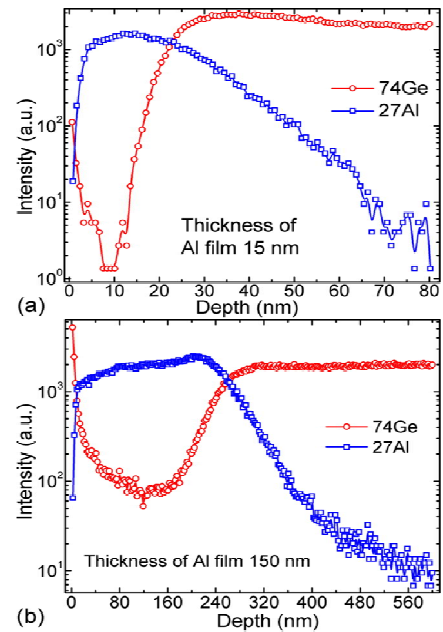

Fig. 3. Depth profiles of Ge and Al in annealed Al-Ge samples for Al film thickness (a) 15 nm and (b) 150 nm.

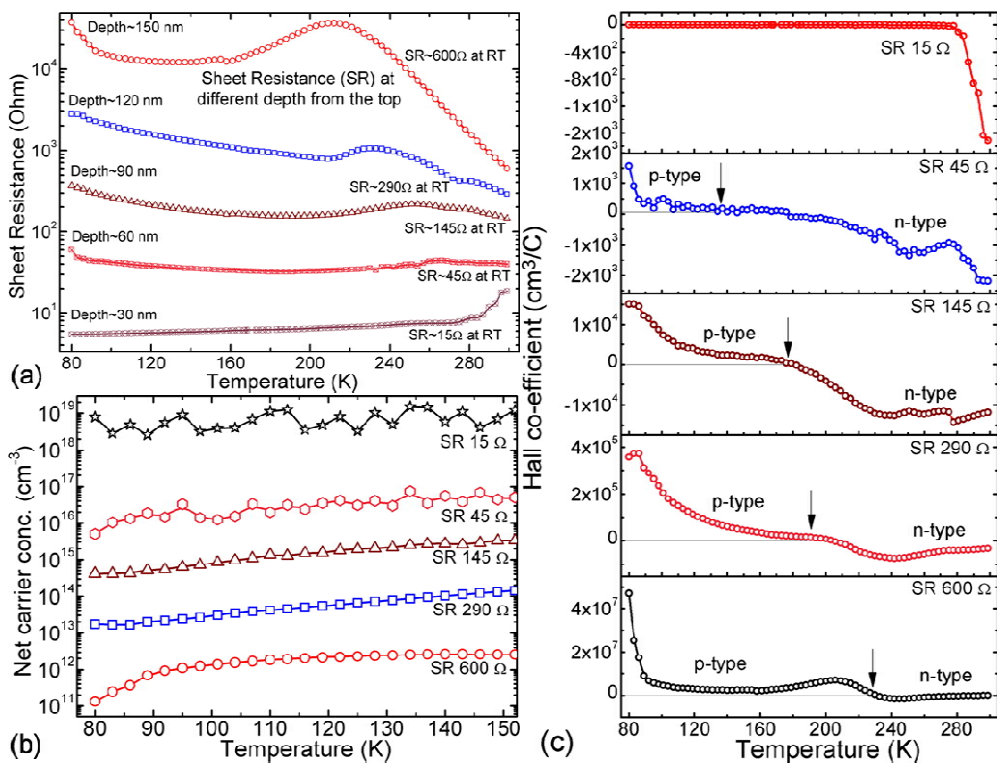

Fig. 4. (a) Sheet resistance (SR) of annealed Al-Ge film at different depths from the top surface. (b) Temperature dependent net carrier concentration and (c) hall co-efficient for different sheet resistances along the depth.

Net carrier concentrations for different sheet resistances [Fig. 4(b)] substantiate the above observation. For higher sheet resistance at higher depth, net carrier concentration decreases as expected. For 15 ohm sheet resistance, corresponding carrier concentration is found to be $\sim 10^{19}$/cm$^3$. It gradually decreases with the depth of the Al-Ge layer and go down up to $10^{11}$/cm$^3$ for sheet resistance $\sim 600$ ohm. At higher temperature, electrical conduction is dominated by thermally generated intrinsic carriers and Al doped Ge layer exhibits n-type conductivity. Nominally doped (conc. below $10^{14}$/cc) p-type Ge shows extrinsic to intrinsic transition near room temperature (220K-260K) and the type of conductivity changes from p-type to n-type (discussed below). For clear comparison, net carrier concentrations are shown in logarithmic scale below 150K avoiding the said extrinsic to intrinsic transition.

The temperature dependence of the type of conductivity is ascertained by hall co-efficient measurement at various depths of Al-Ge [Fig. 4(c)]. The positive hall co-efficient at 80K is found to be $\sim 10^8$ for impurity concentration $10^{11}$/cc. It reduces to almost zero near the top surface (sheet resistance 15 ohm) due to the presence of excessive Al and type of conductivity cannot be ascertained. Hall co-efficient gradually increases at higher depth with higher values of sheet resistance. It is found that, Al-Ge is p-type in extrinsic region and changes the type of conductivity at higher temperatures [indicated in Fig. 4(c)]. This extrinsic to intrinsic transition appears at higher temperature for reduced level of Al and can be explained as the interplay between thermally generated carriers and impurity with the variation of temperature. As the temperature increases, concentration of thermally generated electron and holes increases. Since electrons have a higher mobility than holes, electrical conduction is dominated by electrons and the Al-Ge layer exhibits n-type conductivity at higher temperatures. At low temperatures, impurity conduction plays a dominant role due to reduction in thermally generated charge carriers. It should be noted that Ge crystal used for making contacts exhibit p-type conductivity that changes to n-type one above 170K. Minimum level of Al doping of around $10^{11}$/cc, increases the transition temperature to 230 K. Further increase in Al concentration reduces the transition temperature unlike popular p-type dopant boron that shifts the transition towards higher temperature for its increased concentration [19].

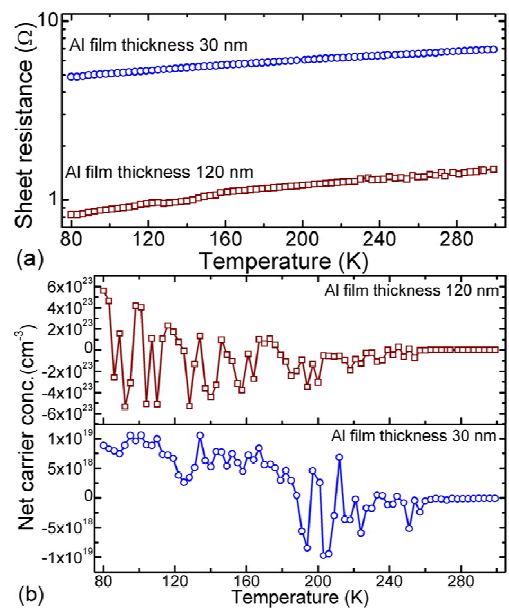

Fig. 5. Sheet resistance (a) and net carrier concentration (b) at the top Al-Ge surface without lapping having different as deposited Al thickness.

The sheet resistance at the top surface is low even for thinner Al-Ge layer although there is a small variation with the thickness of Al layer used for regrowth. Top surface of Al-Ge grown from 30 nm and 120 nm thick Al film exhibit room temperature sheet resistance values around 7 ohm and 1.5 ohm respectively [Fig. 5(a)]. Whereas sheet resistance of Al-Ge (Al thickness~120nm), at depths 120 nm and 30 nm are found to be 290 ohm and 15 ohm respectively. This observation emphasizes that irrespective of the deposited Al thickness, the top Al-Ge surface contains high level of Al impurity and exhibits metallic characteristic which is desired for making electrical contacts. As shown in Fig. 5 (b), the Al concentration is as high as $10^{23}$/cc and $10^{19}$/cc on the top surface of regrown Al-Ge from Al film thickness 120 nm and 30nm respectively. Al concentration $>10^{20}$/cm$^3$ found near the surface is comparable to the electrically active impurity concentration $\sim 10^{20}$/cm$^3$ achieved by Al implantation [20].

IV. CONCLUSION

Rugged p$^+$ electrical contact on p-type Ge crystal has been fabricated by solid state regrowth of thermally deposited Al on Ge. Sufficiently large numbers of samples with wide range of Al film thickness are processed and analyzed by SIMS to find the diffusion length of Al and Ge across the interface. It is established that thickness of the regrown Al-Ge contact can be controlled by varying the thickness of deposited Al film. Al doped Ge layer exhibits p-type conductivity in the extrinsic range of temperature. Dopant concentration at the top surface of Al-Ge on p-type Ge can be as high as $10^{20}$/cm$^3$ that gradually decreases along the depth resulting in p+/p type of arrangement.

ACKNOWLEDGMENT

Authors are thankful to Dr. T. V. C. Rao for his constant encouragement and support. Unconditional help received from all members of Crystal Technology Section is gratefully acknowledged.